\documentstyle[12pt]{article}

\def\F{{\cal F}}
\def\N{{\cal N}}
\def\D{\partial}
\def\k{{\bf k}}
\def\a{{\bf a}}

\begin{document}

\begin{titlepage}

\begin{flushright}
UCLA/01/TEP/8 \\
May, 2001
\end{flushright}
\bigskip

\begin {center}

{\LARGE 
Prepotential Recursion Relations 
in $\N$=2 Super Yang-Mills with  
Adjoint Matter
\\\ }
\\ [15mm]

Gordon Chan\footnote{E-mail:  {\tt morgoth@physics.ucla.edu}}
\\ [12mm]
{\it Department of Physics and Astronomy \\
\medskip University of California \\
\medskip Los Angeles, CA 90095, USA} \\ [15mm]

\end{center}

\begin{abstract}

Linear recursion relations for the instanton corrections to the
effective prepotential are derived for $\N$=2 supersymmetric gauge theories 
with one hypermultiplet in the adjoint representation of $SU(N)$ using
the Calogero-Moser parameterization of the Seiberg-Witten spectral curves.
S-duality properties of the Calogero-Moser parameterization
and conjectures on the Seiberg-Witten spectral curves 
generalized to arbitrary simply laced
classical gauge groups are also discussed.

\end{abstract}

\end{titlepage}

\setcounter{section}{0}
\setcounter{subsection}{0}
\setcounter{equation}{0}
\setcounter{footnote}{0}

%
\addtocounter{section}{1}
{\large {\bf \thesection. Introduction}}
\medskip

Over the past half decade there has been great progress in understanding
non-perturbative dynamics of $\N$=2 SUSY gauge theories 
\cite{SW1}-\cite{ArgShap}.  
Non-perturbative corrections in the weak coupling corresponding 
to instanton effects \cite{Seiberg} were evaluated by field theory methods
\cite{ShiftVain}-\cite{Dor} and various other ways via the 
Seiberg-Witten ansatz by calculating the period integrals corresponding to
the quantum moduli parameters representing the set of vacuum expectation
values of the Higgs fields \cite{Lerche}-\cite{DPK}.
In a previous paper \cite{Recur}, a linear set of recursion relations
for the instanton corrections to the effective prepotential $\F$
was found for a class of $\N$=2 SUSY Yang-Mills theories with hypermultiplets 
in the fundamental representation of a classical gauge group ${\cal G}$,
which reproduced the results of previous recursion relations 
\cite{Matone}\cite{Edel}
and other methods \cite{Lerche}-\cite{DPK}.

Connections between Seiberg-Witten theory and integrable systems was
first made for the case of pure $\N$=2 super Yang-Mills theory
in connection with Toda lattices \cite{MarWar}
and Whitham theory \cite{Whitham}.  Later connections between
$\N$=2 super Yang-Mills theory with one hypermultiplet in the adjoint
representation of the gauge group, and the Hitchin \cite{Donagi}
and Calogero-Moser \cite{Martinec} integrable systems was made.  (There are
claims that the Calogero-Moser integrable system can be derived from 
the Hitchin integrable system \cite{Markman}).  Convenient parameterizations
of the Calogero-Moser integrable system useful for performing explicit
Seiberg-Witten type of calculations were discovered \cite{Calegro} and
forms the starting point of the present paper.

The Calogero-Moser construction \cite{Calegro} of the Seiberg-Witten 
solution for $\N$=2 super Yang-Mills theory with
one hypermultiplet in the adjoint representation of the gauge
group $SU(N)$ and the renormalization group like
equation for the prepotential $\F$, led us to the discovery of a general
recursion relation expressing the n-th order instanton correction to the
prepotential $\F$ in terms of the (n-1)-th, ..., first order instanton
corrections.

We start off by reviewing the Calogero-Moser construction of the 
Seiberg-Witten solution for $\N$=2 
super Yang-Mills theory with one hypermultiplet in the adjoint
representation of the gauge group $SU(N)$.  The renormalization group
type equation for the prepotential $\F$ is discussed next and it is
shown how it can be used to determine the instanton corrections
to the prepotential to arbitrary order in an efficient manner.  Recursion
relations for the instanton corrections are then derived and compared
with previous results.  
S-duality properties of the Calogero-Moser construction of the Seiberg-Witten
solution are used to discuss the "dual" prepotential $\F_D$ of the dual
magnetic sector of the theory.
Conjectures for possible Seiberg-Witten spectral curves
for simply laced cases of other classical gauge groups are then discussed.

\vspace{7mm}
\addtocounter{section}{1}
{\large {\bf \thesection.  The Seiberg-Witten Solution for Super Yang-Mills with One Adjoint Hypermultiplet}}
\medskip

The Seiberg-Witten (SW) ansatz gives a prescription for determining the
prepotential of the effective action for $\N$=2 supersymmetric Yang-Mills
gauge theories, as well as for determining the spectrum of BPS states.

For supersymmetric Yang-Mills theories with an asymptotic free coupling 
and one adjoint hypermultiplet in the adjoint representation of a classical
gauge group, general arguments based on the holomorphicity of $\F$, 
perturbative non-renormalization theorems beyond 1-loop order, the nature
of instanton corrections, and restrictions of $U(1)_R$ invariance constrain
$\F$ to have the form
\begin{eqnarray}
\F(a) & = & \frac{\tau}{2} \sum_{i=1}^r a_i^2 -\frac{1}{8\pi i} 
\sum_{\alpha \in {\cal R(G)} } \left\{(\alpha \cdot a)^2 log (\alpha \cdot a)^2 
\right.
\nonumber \\
& & \left. - (\alpha \cdot a+m)^2 log(\alpha \cdot a+m)^2\right\} 
+ \sum_{n=1}^\infty \frac{q^n}{2\pi n i}\F^{(n)}(a) 
\label{eq:pertpre}
\end{eqnarray}
where $\alpha$ are the roots of the gauge group $\cal{G}$.  
For $SU(N)$, the traceless condition $\sum_{i=1}^N a_i = 0$ is imposed.

The SW ansatz for determining the full prepotential $\F$ is based on 
a choice of a fibration of spectral curves over the space of vacua,
and of a meromorphic 1-form $d\lambda$ on each of these curves.
The renormalized order parameters $a_k$ of the theory, their duals
$a_{D,k}$, and the prepotential $\F$ are given by
\begin{eqnarray}
2\pi i a_k = \oint_{A_k} d \lambda, &
\; 2\pi i a_{D, k} = {\displaystyle \oint_{B_k} } d \lambda, &
\; a_{D,k} = \frac{\D \F}{\D a_k}
\label{eq:sw}
\end{eqnarray}
with $A_k, B_k$ a suitable set of homology cycles on the spectral curves.

For $\N=2$ supersymmetric gauge theories with gauge group $SU(N)$ and one
hypermultiplet in the adjoint representation, a
convenient parameterization for the spectral curves and meromorphic 1-forms 
is the Calogero-Moser case of \cite{Calegro}
\begin{eqnarray}
f(k-\frac{m}{2}, z) = 0, & d \lambda = k dz \label{eq:spectral}
\end{eqnarray}
where
\begin{eqnarray}
f(k,z)  & = & {\displaystyle
\frac{1}{\vartheta_1 (\frac{z}{2 \omega_1} | \tau)}\sum_{n=0}^N
\frac{1}{n!} \frac{\D^n}{\D z^n} \vartheta_1 (\frac{z}{2 \omega_1} | \tau)
(-m \frac{\D}{\D k})^n H(k | \k) } \\ \label{eq:tspec}
H(x | \k) & = & \prod^N_{j=1}(x-k_j) \equiv (x-k_i) H_i(x | \k) \\
\vartheta_1 (z | \tau) & = & \sum_{n \in Z} (-1)^n q^{(n+1/2)^2/2} 
e^{2\pi i(2n+1)z} \label{eq:theta}
\end{eqnarray}
along with a corresponding basis of 
$A_k, B_k$ homology cycles as described in \cite{Calegro}.
This particular choice of parameterization for the spectral curves has 
the geometry of a foliation over a base torus $\Sigma$, where the complex
modulus $\tau$ of the torus $\Sigma$ is related to the gauge coupling
$g$ and the $\theta$-angle of the gauge theory by
\begin{eqnarray}
\tau = \frac{\theta}{2\pi} + \frac{4\pi i}{g^2}
\label{eq:torus}
\end{eqnarray}

Substituting (\ref{eq:theta}) into
(\ref{eq:spectral}) produces a
simplified form for the spectral curves
\begin{eqnarray}
\sum_{n \in Z} (-1)^n q^{\frac{1}{2}n(n-1)}e^{nz} H(k-mn | \k) = 0
\label{eq:simspec}
\end{eqnarray}

Substituting (\ref{eq:simspec}) into (\ref{eq:sw}) using the 
Calogero-Moser parameterization (\ref{eq:spectral}) for the 1-form and
performing a weak coupling expansion in powers of $q$ similar to 
the methods in \cite{Calegro}, the integral for the
quantum order parameters $a_i$'s in terms of the classical
order parameters $k_i$'s were calculated order by order in $q$
producing a simplified expression of
\begin{eqnarray}
a_i &  = &  k_i + \sum_{n=1}^\infty q^n \Delta^{(n)}_i (k)
\nonumber
\end{eqnarray}
where
\begin{eqnarray}
\sum_{n=1}^\infty q^n \Delta^{(n)}_i (k) & = & \sum_{j=2}^\infty
\;
\sum^\infty_{\stackrel{\alpha_1, \cdots, \alpha_j = -\infty, \neq 0}
{\alpha_1 + \cdots + \alpha_j = 0} }
\frac{(-1)^j}{j!} (\frac{\D}{\D k_i})^{j-1} 
\prod_{l=1}^j \left[\frac{H(k_i - \alpha_l m | \k)}{H_i (k_i | \k)} q^{\alpha_l^2/2}
\right] \nonumber \\ \label{eq:qorder}
\end{eqnarray}

The first few $\Delta_i$'s are

\begin{eqnarray}
\Delta_i^{(1)}(a) & = &
\frac{\D}{\D a_i}
\left[ \frac{H(a_i-m | \a)H(a_i+m | \a)}{H_i (a_i | \a)^2} \right]
\nonumber \\
\Delta_i^{(2)}(a) & = &
\frac{1}{4}
\frac{\D^3}{\D a_i^3} 
\left[\frac{H(a_i-m | \a)H(a_i+m | \a)}{H_i (a_i | \a)^2}\right]^2
\nonumber \\
\Delta_i^{(3)}(a) & = &
\frac{1}{36} \frac{\D^5}{\D a_i^5} 
\left[\frac{H(a_i-m | \a)H(a_i+m | \a)}{H_i (a_i | \a)^2}\right]^3
\nonumber \\
& - & \frac{1}{2}
\frac{\D^2}{\D a_i^2}
\left[ \frac{H(a_i-2m| \a) H(a_i+m | \a)^2}{H_i (a_i | \a)^3} \right]
\nonumber \\
&-& \frac{1}{2}
\frac{\D^2}{\D a_i^2}
\left[ \frac{H(a_i-m | \a)^2 H(a_i+2m | \a)}{H_i (a_i | \a)^3} \right]
\nonumber \\
\Delta_i^{(4)}(a) & = & 
\frac{1}{576} \frac{\D^7}{\D a_i^7} 
\left[\frac{H(a_i-m | \a)H(a_i+m | \a)}{H_i (a_i | \a)^2}\right]^4
\nonumber \\
&+& \frac{\D}{\D a_i} 
\left[ \frac{H(a_i-2m | \a) H(a_i+2m | \a)}{H_i (a_i | \a)^2} \right]
\nonumber \\
& - & \frac{1}{6}
\frac{\D^4}{\D a_i^4}
\left[\frac{H(a_i-2m | \a) H(a_i-m | \a) H(a_i+m | \a)^3}{H_i (a_i | \a)^5}\right] 
\nonumber \\
& - & \frac{1}{6}
\frac{\D^4}{\D a_i^4}
\left[\frac{H(a_i-m | \a)^3 H(a_i+m | \a)H(a_i+2m | \a)}{H_i (a_i | \a)^5}\right]
\nonumber 
\end{eqnarray}

This result can be derived in a more transparent manner by rewriting 
the spectral curves (\ref{eq:simspec}) as
\begin{eqnarray}
k & \equiv & k_i + F_i (k)
\end{eqnarray}
where
\begin{eqnarray}
F_i (k) & = &   \sum_{n \in Z, \neq 0} (-1)^{n+1}
q^{\frac{1}{2} n(n-1)} e^{nz} \frac{H(k-nm | \k)}{H_i (k | \k)}
\end{eqnarray}
An iterative solution expanded around $k = k_i$ to all orders 
in small q is given by
\begin{eqnarray}
k = k_i + \sum_{n=1}^\infty y_n, & &  
y_n = \frac{1}{n!} \frac{\D^{n-1} }{\D k^{n-1} } F_i^n (k) |_{k= k_i}
\label{eq:iterative}
\end{eqnarray}

In a similar manner as in \cite{Calegro}, $z$ is substituted 
with $w =  e^z$ in the the SW differential (\ref{eq:spectral}),
\begin{eqnarray}
dz = \frac{dw}{w}, && 
d \lambda = k dz  = k \frac{dw}{w} 
\label{eq:subswdiff}
\end{eqnarray}
along with the iterative solution (\ref{eq:iterative}).  
Performing the integral around the
appropriate $A_i$ cycle corresponding to $k_i$ as prescribed in
\cite{Calegro}, reproduces (\ref{eq:qorder}) \cite{Pdhoker}.

\vspace{7mm}
\addtocounter{section}{1}
{\large {\bf \thesection.  Renormalization Group Type Equations}}
\medskip

In \cite{Calegro}, a renormalization group type equation for the   
prepotential $\F$ was derived
\begin{equation}
{\D\F\over\D\tau} = {1\over{4\pi i}} \sum_{j=1}^r \oint_{A_j} k^2 dz
\label{eq:rg}
\end{equation}
up to an additive term independent of $a_i$ and $k_i$ which is 
physically immaterial.

Substituting the $SU(N)$ spectral curves (\ref{eq:simspec}) 
into (\ref{eq:rg}) using the Calogero-Moser
parameterization (\ref{eq:spectral}) for the 1-form and solving
the integral in the weak coupling limit of small $q$
gives the renormalization group like equation for the prepotential
$\F$ in terms of the classical order parameters $k_i$'s

\begin{eqnarray}
{\D\F\over\D\tau} & = &  \frac{1}{2} \sum_{i=1}^r k_i^2
+ \sum_{i=1}^r \sum_{n=1}^\infty q^n k_i \Delta_i^{(n)}(k)
+ \sum_{i=1}^r \sum_{n=1}^\infty q^n \Omega_i^{(n)}(k)
\nonumber
\end{eqnarray}
where
\begin{eqnarray}
\sum_{n=1}^\infty q^n \Omega^{(n)}_i (k) & = & \sum_{j=2}^\infty
\;
\sum^\infty_{\stackrel{\alpha_1, \cdots, \alpha_j = -\infty, \neq 0}
{\alpha_1 + \cdots + \alpha_j = 0} }
\frac{(-1)^j}{j(j-2)!} (\frac{\D}{\D k_i})^{j-2} 
\prod_{l=1}^j \left[ \frac{H(k_i - \alpha_l m | \k)}{H_i (k_i | \k)} q^{\alpha_l^2/2}
\right] \nonumber \\
& &
\label{eq:renormg}
\end{eqnarray}

The first few $\Omega_i$'s are

\begin{eqnarray}
\Omega_i^{(1)}(a) & = &
\frac{H(a_i-m | \a)H(a_i+m | \a)}{H_i (a_i | \a)^2}
\nonumber \\
\Omega_i^{(2)}(a) & = &
\frac{3}{4} \frac{\D^2}{\D a_i^2} 
\left[\frac{H(a_i-m | \a)H(a_i+m | \a)}{H_i (a_i | \a)^2}\right]^2
\nonumber \\
\Omega_i^{(3)}(a) & = &
\frac{5}{36} \frac{\D^4}{\D a_i^4} 
\left[\frac{H(a_i-m | \a)H(a_i+m | \a)}{H_i (a_i | \a)^2}\right]^3
\nonumber \\
&-& \frac{\D}{\D a_i}
\left[ \frac{H(a_i-2m| \a) H(a_i+m | \a)^2}{H_i (a_i | \a)^3} \right]
\nonumber \\
&-& \frac{\D}{\D a_i}
\left [\frac{H(a_i-m | \a)^2 H(a_i+2m | \a)}{H_i (a_i | \a)^3} \right]
\nonumber \\
\Omega_i^{(4)}(a) & = & 
\frac{7}{576} \frac{\D^6}{\D a_i^6} 
\left[\frac{H(a_i-m | \a)H(a_i+m | \a)}{H_i (a_i | \a)^2}\right]^4
\nonumber \\
&+& \frac{H(a_i-2m | \a) H(a_i+2m | \a)}{H_i (a_i | \a)^2}
\nonumber \\
& - & \frac{2}{3}
\frac{\D^3}{\D a_i^3}
\left[\frac{H(a_i-2m | \a) H(a_i-m | \a) H(a_i+m | \a)^3}{H_i (a_i | \a)^5}\right] 
\nonumber \\
& - & \frac{2}{3}
\frac{\D^3}{\D a_i^3}
\left[\frac{H(a_i-m | \a)^3 H(a_i+m | \a)H(a_i+2m | \a)}{H_i (a_i | \a)^5}\right]
\nonumber 
\end{eqnarray}

%
%
\vspace{7mm}
\addtocounter{section}{1}
{\large {\bf \thesection. Recursion Relations for the Prepotential $\F$}}
\medskip

In \cite{Recur}, an efficient algorithm for deriving a set of recursion
relations for the instanton corrections was discovered.  Using similar
methods as \cite{Recur}, a similar set of recursion relations for
the prepotential $\F$ was determined.

A very direct way of deriving the form of the instanton corrections to the
prepotential $\F$ involves
substituting (\ref{eq:pertpre}) and (\ref{eq:qorder}) into 
(\ref{eq:renormg}) to get
\begin{eqnarray}
\sum_{n=1}^\infty q^n \F^{(n)}(a) & = & 
\sum_{i=1}^r \sum_{n=1}^\infty q^n \Omega^{(n)}_i (k) 
- \frac{1}{2}
\sum_{i=1}^r \left[\sum_{n=1}^\infty q^n \Delta^{(n)}_i (k)\right]^2
\label{eq:istart}
\end{eqnarray}
Then (\ref{eq:qorder}) is substituted into (\ref{eq:istart})
and expanded in powers of $q$, replacing the $k_i$'s with $a_i$'s.
The n-th order instanton correction to the prepotential $\F$ 
finally takes on the form
\begin{eqnarray}
\F^{(n)}(a) & = & \sum_{i=1}^r \Omega^{(n)}_i (a) 
-\frac{1}{2} \sum_{i=1}^r \sum_{\stackrel{j,l=1}{j+l=n}}^n 
\Delta^{(j)}_i(a) \Delta^{(l)}_i(a)
\nonumber \\
& - & \sum_{i=1}^{n-1} \frac{1}{i!}
\sum_{\stackrel{\beta_1, \cdots, \beta_{i+1} = 1}{\beta_1 + \cdots + \beta_{i+1} = n} }^{i-1}
\sum_{\alpha_1, \cdots, \alpha_i =1}^r 
\left[\prod_{j=1}^i \Delta_{\alpha_j}^{(\beta_j)}(a)\right]
\left(\prod_{l=1}^i \frac{\D}{\D a_{\alpha_l} }\right) \F^{(\beta_{i+1})}(a)
\nonumber \\ \label{eq:prepot}
\end{eqnarray}

The first few $\F^{(n)}(a)$'s are

\begin{eqnarray}
\F^{(1)}(a) & = & \sum_{i=1}^r \Omega_i^{(1)}(a)
\nonumber \\
\F^{(2)}(a) & = & \sum_{i=1}^r \Omega_i^{(2)}(a)
-\frac{1}{2} \sum_{i=1}^r (\Delta_i^{(1)}(a))^2
-\sum_{i=1}^r \Delta_i^{(1)}(a) \frac{\D \F^{(1)}(a)}{\D a_i}
\nonumber \\
\F^{(3)}(a) & = & \sum_{i=1}^r \Omega_i^{(3)}(a)
-\frac{1}{2} \sum_{i=1}^r \left[2\Delta_i^{(1)}(a) \Delta_i^{(2)}(a)\right]
\nonumber \\
& - & \sum_{i=1}^r \left[
\Delta_i^{(1)}(a) \frac{\D \F^{(2)}(a)}{\D a_i}
+ \Delta_i^{(2)}(a) \frac{\D \F^{(1)}(a)}{\D a_i} \right]
\nonumber \\
&-& \frac{1}{2!} \sum_{i,j=1}^r 
\Delta_i^{(1)}(a) \Delta_j^{(1)}(a) \frac{\D^2 \F^{(1)}(a)}{\D a_i \D a_j}
\nonumber \\
\F^{(4)}(a) & = & \sum_{i=1}^r \Omega_i^{(4)}(a)
-\frac{1}{2} \sum_{i=1}^r \left[2\Delta_i^{(1)}(a) \Delta_i^{(3)}(a)
+(\Delta_i^{(2)}(a))^2 \right]
\nonumber \\
& - & \sum_{i=1}^r \left[
\Delta_i^{(1)}(a) \frac{\D \F^{(3)}(a)}{\D a_i}
+ \Delta_i^{(2)}(a) \frac{\D \F^{(2)}(a)}{\D a_i}
+ \Delta_i^{(3)}(a) \frac{\D \F^{(1)}(a)}{\D a_i} \right]
\nonumber \\
&-& \frac{1}{2!} \sum_{i,j=1}^r \left[
\Delta_i^{(1)}(a) \Delta_j^{(1)}(a) \frac{\D^2 \F^{(2)}(a)}{\D a_i \D a_j}
+ 2\Delta_i^{(1)}(a) \Delta_j^{(2)}(a) \frac{\D^2 \F^{(1)}(a)}{\D a_i \D a_j}
\right] \nonumber \\
&-& \frac{1}{3!} \sum_{i,j,k=1}^r 
\Delta_i^{(1)}(a) \Delta_j^{(1)}(a) \Delta_k^{(1)}(a)
\frac{\D^3 \F^{(1)}(a)}{\D a_i \D a_j \D a_k}
\end{eqnarray}

\vspace{7mm}
\addtocounter{section}{1}
{\large {\bf \thesection.  Comparison With Previous Results}}
\medskip

In the limit the full hypermultiplet is decoupled with
$\tau \rightarrow \infty, m \rightarrow \infty$ while keeping
constant the parameters $k_i$ and $\Lambda$:
\begin{eqnarray}
\Lambda^{2N} = (-1)^N m^{2N} q & &  q = e^{2\pi i \tau}
\end{eqnarray} 
equations (\ref{eq:qorder}) and (\ref{eq:renormg}) break down to their
corresponding equations in the pure $SU(N)$ gauge theory cases 
\cite{DPK}\cite{Recur}\cite{Renorm}.

In the $\N$=4 limit where $m \rightarrow 0$, all the $\Omega_i$ and
$\Delta_i$ terms in (\ref{eq:qorder}) and (\ref{eq:renormg}) vanish
and reproduces the expected prepotential
\begin{eqnarray}
\F (a) = \frac{\tau}{2}\sum_{i=1}^N a_i^2
\end{eqnarray}
for $SU(N)$.

For $SU(2)$, the existing results in the literature have the instanton
expressed in term of
\begin{eqnarray}
a_1 = a, & a_2 = -a
\end{eqnarray}
Explicit evaluations of the first three instanton corrections 
are

\begin{eqnarray}
\F^{(1)}(a) & = & \frac{m^4}{2a^2} \nonumber \\
\frac{\F^{(2)}(a)}{2} & = & -\frac{9 m^6}{16 a^4} 
+ \frac{5 m^8}{64 a^6} \nonumber \\
\frac{\F^{(3)}(a)}{3} & = & \frac{m^6}{a^4} + \frac{25 m^8}{48 a^6}
-\frac{67 m^{10}}{192 a^8} + \frac{3 m^{12}}{64 a^{10}} 
\label{eq:su2result}
\end{eqnarray}
which disagree with results in the literature beyond one instanton
\cite{Minahan}
\footnote{
In the limit of decoupling
the full $SU(2)$ adjoint hypermultiplet in \cite{Minahan}, 
there is a discrephency of a factor
$\frac{1}{2}$ with the pure $SU(2)$ results of \cite{Lerche}\cite{Recur}.}
, but agrees in the limit where the full
hypermultiplet decouples \cite{Recur}.  It turns out that peforming the 
Seiberg-Witten elliptic function calculation in \cite{Dorey} 
to higher instanton orders reproduces the
instanton calculations of \cite{Minahan}.  

On the other hand, the $SU(2)$ spectral curve from the 
Calogero-Moser construction (\ref{eq:spectral}) can be explicitly 
shown to be equivalent to the $SU(2)$ mass deformed $\N$=4 
spectral curve construction \cite{SW2} up to reparameterizations
of the classical order parameters $k_i$'s.  This spectral curve
forms a crucial part of the elliptic function calculation in 
\cite{Dorey}.

One possible problem with the elliptic function calculation in \cite{Dorey}
is the assumption of Matone's relation 
\cite{Matone}\cite{Sonnen} holding in the 
presence of an adjoint hypermultiplet.  Generalizations of Matone's relation
for classes of $\N$=2 SUSY gauge theories with fundamental hypermultiplets 
was proven in general \cite{Renorm} and corresponds to a renormalization
group type of equation for the prepotential $\F$.  For the case of one adjoint
hypermultiplet, a renormalization group equation for the prepotential $\F$ 
was derived \cite{Calegro} and calculated to all orders (\ref{eq:renormg})
which differs greatly from Matone's relation and 
\cite{Sonnen}\cite{Renorm}, but
agrees with the latter cases in the limit the full adjoint hypermultiplet is
decoupled \cite{Calegro}.

\vspace{7mm}
\addtocounter{section}{1}
{\large {\bf \thesection. S-Duality Properties}}
\medskip

A closer examination of the Calogero-Moser parameterization of the
Seiberg-Witten spectral curves and 1-form (\ref{eq:spectral}) and
(\ref{eq:tspec}) reveals there's an implicit S-duality present.  

Using the transformation property of the theta functions
\begin{eqnarray}
\vartheta_1 \left(\frac{z}{\tau} | -\frac{1}{\tau} \right)
= \sqrt{-i\tau} \exp \left( \frac{iz^2}{\pi\tau} \right) 
\vartheta_1 (z | \tau)
\end{eqnarray}  
and substituting it into (\ref{eq:spectral}) and (\ref{eq:tspec}) shows
explicitly that the form of the spectral curves and 1-form are indeed invariant
under S-duality transformations up to reparameterizations of the
classical order parameters $k_i$'s.  Correspondingly, the roles of the
A and B cycles in the Seiberg-Witten ansatz (\ref{eq:sw}) are interchanged
under S-duality transformations.  

With this explicit S-duality, the corresponding weakly coupled "dual" 
theory in the magnetic sector of the theory 
expanded around a small "dual" coupling constant
\begin{eqnarray}
q_D = e^{2\pi i \tau_D} & \tau_D = -\frac{1}{\tau}
& \tau \rightarrow i0^+ 
\end{eqnarray}
will have a corresponding "dual" prepotential $\F_D(a_D)$ identical in
form to the prepotential $\F(a)$ in 
(\ref{eq:pertpre}) and (\ref{eq:prepot}) with the
corresponding substitutions of the coupling constant and 
quantum order parameters to their "dual" counterparts
\begin{eqnarray}
q \rightarrow q_D & & a_i \rightarrow a_{D,i} 
\end{eqnarray}
respectively.  This can be interpreted as a non-perturbative expansion
of the theory, where the dynamics of the strongly coupled
regime in the electric
sector of the theory is described by the dynamics of a corresponding weakly 
coupled "dual" theory in the magnetic sector of the same theory.  (Other
strong coupling expansions in the same spirit were performed in 
\cite{Lerche}\cite{dhokerstrong}\cite{EdelStrong}).

Considering there are claims that the Calogero-Moser system 
can be constructed explicitly from the Hitchin system \cite{Markman}, this
S-duality is like a realization of the Donagi-Witten construction of
Seiberg-Witten theory using the Hitchin system
\cite{Donagi} where an underlying S-duality and general $SL(2, Z)$
symmetry is built 
into the geometry of the foliation over a base torus $\Sigma$
construction (\ref{eq:torus}) from the start.
In the prepotential calculations performed around small 
coupling $q$ or $q_D$, the S-duality is explicitly broken while the
underlying spectral curve (\ref{eq:spectral}) is invariant 
under S-duality and in general an $SL(2, Z)$ symmetry 
\cite{SW2}\cite{Donagi}.

\vspace{7mm}
\addtocounter{section}{1}
{\large {\bf \thesection. Generalizations to Other Gauge Groups}}
\medskip

Generalizations of the $SU(N)$ Calogero-Moser integrable system were
investigated in \cite{DHokerPhong} for various cases of twisted and untwisted
gauge groups, but stopped short of producing parameterizations suitable
for use as Seiberg-Witten spectral curves.
Possible parameterizations to general untwisted classical gauge groups
can be conjectured starting from the $SU(N)$ spectral curves and placing
appropriate constraints such that decouplings of the full adjoint
hypermultiplet reproduce the pure gauge theory results 
\cite{Brandhuber}\cite{Daniel}\cite{Hanany}\cite{DHokerGen}.

In the spirit of \cite{Hanany}\cite{DHokerGen}, one possibility is to 
replace the $H(k)$ polynomial with
\begin{eqnarray}
H(x | \k ) &\rightarrow &
H(x | \k)  =  \prod^N_{j=1}(x^2-k^2_j) \equiv (x-k_i)(x+k_i) H_i(x | \k) 
\end{eqnarray}
in the Calogero-Moser parameterization of the SW spectral curves 
(\ref{eq:simspec}).

The appropriate limits for full hypermultiplet decoupling are 
$\tau \rightarrow \infty, m \rightarrow \infty$ while keeping
constant the parameters $k_i$ and $\Lambda$:
\begin{eqnarray}
SO(2r) & & \Lambda^{4r-4} \equiv m^{4r-4}q \nonumber \\
SO(2r+1) & & \Lambda^{4r-2} \equiv m^{4r-2}q \nonumber \\
SO(2r) & & \Lambda^{4r+4} \equiv m^{4r+4}q 
\end{eqnarray}
where $q=e^{2\pi i \tau}$.

\vspace{7mm}
\addtocounter{section}{1}
{\large {\bf \thesection.  Acknowledgments}}
\medskip

	The author is very grateful to Eric D'Hoker for guidance.  The
author would also like to thank Anton Ryzhov, Nick Dorey, Alex Buchel,
and Gordon Chalmers for various discussions. This
research was financed partially by NSERC.

%
%

\end{document}